\documentclass[intlimits,twoside,a4paper]{article}

\usepackage{amsmath,amssymb}
\usepackage{graphicx}
\usepackage{wrapfig}

\usepackage[T2A]{fontenc}
\usepackage[cp1251]{inputenc}

\usepackage[eqsecnum]{cmpj2}

\issue{2013}{16}{1}{13002}

\doinumber{10.5488/CMP.16.13002}

\title[How to solve Fokker-Planck equation?]%
{How to solve Fokker-Planck equation treating mixed eigenvalue spectrum?%
}
\author[M. Brics, J. Kaupu\v{z}s, R. Mahnke]{M. Brics\refaddr{label1}, J. Kaupu\v{z}s\refaddr{label2},
R. Mahnke\refaddr{label1}}
\addresses{
\addr{label1}  Institute of Physics, Rostock University, D--18051 Rostock, Germany
\addr{label2}  Institute of Mathematics and Computer Science, University of Latvia,
	       LV--1459 Riga, Latvia
}

\date{Received  July 3, 2012, in final form August 24, 2012}
\authorcopyright{M. Brics, J. Kaupu\v{z}s, R. Mahnke, 2013}

\begin{document}

\maketitle

\begin{abstract}
An analogy of the Fokker-Planck equation (FPE) with the Schr\"o\-dinger equation
allows us to use quantum mechanics technique  to find the analytical solution of the FPE
in a number of cases. However, previous studies have been limited to the
Schr\"odinger potential with a discrete eigenvalue spectrum. Here, we will show how this
approach can be  also applied to a mixed eigenvalue spectrum with bounded and free states.
We solve the FPE with boundaries located at $x=\pm L/2$ and take the limit $L\rightarrow\infty$,
considering the examples with constant Schr\"{o}dinger potential and with P\"{o}schl-Teller potential.
An oversimplified approach was proposed earlier  by M.T.~Araujo and E.~Drigo Filho.
A detailed investigation of the two examples shows that the correct solution,
obtained in this paper, is consistent with the expected Fokker-Planck dynamics.
\pacs 05.10.Gg
\end{abstract}

\section{Introduction}

The one--dimensional Fokker-Planck equation (FPE) for the probability density 
$p(x,t)$, depending on variable $x$ and time $t$, assumes the generic
form~\cite{Smoluchowski06,Smoluchowski04,Risken,Gardiner,Schadschneider,MKLbook,Lo}
\begin{equation}
 \frac{\partial p(x,t)}{\partial t} = - \frac{\partial}{\partial x}
\left[ f(x,t) p(x,t) \right] + \frac{\partial^2}{\partial x^2}
\left[ \frac{D(x,t)}{2} p(x,t) \right] \;.
\label{eq:genfp}
\end{equation}
Here, the drift coefficient or force $f(x,t)$ and
the diffusion coefficient $D(x,t)$ depend on $x$ and $t$ in general.
The Fokker-Planck equation is related to the Smoluchowski equation.
Starting with pioneering works by Marian Smoluchowski~\cite{Smoluchowski06,Smoluchowski04},
these equations have been historically
used to describe the Brownian-like motion of particles. The Smoluchowski equation describes the
high-friction limit, whereas the Fokker-Planck equation refers to the general case.

The FPE provides a very useful tool for modelling a wide variety
of stochastic phenomena arising in physics, chemistry, biology,
finance, traffic flow, etc.~\cite{Risken,Gardiner,Schadschneider,MKLbook}.
Given the importance of the Fokker-Planck equation,
different analytical and numerical methods have been proposed for its solution.
As it is well known, the stationary solution of FPE can be given
in a closed form if the condition of a detailed balance holds.
The study of the time-dependent solution is a much more complicated problem.
The FPE~(\ref{eq:genfp}) with a general time-dependence
and a special $x$-dependence of the drift and diffusion coefficients
has been studied analytically in~\cite{Lo} using Lie algebra.
This method is applicable when the Fokker-Planck equation has a definite
algebraic structure, which makes it possible to employ the Lie algebra and the Wei-Norman theorem.
Generally, there are only a few exactly solvable cases. A simple example is
a system with constant diffusion coefficient and harmonic interaction
of the form $f(x) = - \rd V(x)/ \rd x$ with harmonic potential $V(x) \sim x^2$.
The case with double-well potential is already quite non-trivial and requires
a numerical approach~\cite{ChLiebe}.

The known relation between the Fokker-Planck equation and the Schr\"odinger
equation can also be used. This approach allows us to apply the well known
methods of quantum mechanics. In particular, analytical solutions
can be found in the cases, where the eigenvalues and eigenfunctions for
the considered Schr\"odinger potential are known.
For a general Schr\"odinger potential, numerical
treatments used in quantum mechanics, such as the Crank-Nicolson time
propagation with implicit Numerov's method for second order derivatives~\cite{Bauer},
are very useful.
To apply it to Schr\"odinger-type equation, we just need to replace the real time
step $\Delta t$ by an imaginary time step $\Delta t\rightarrow -\ri \Delta t$.
In quantum mechanics, this is called imaginary time propagation and is used for
calculation of both ground states and excited states.
The analytical studies of mapping the
FPE to Schr\"odinger equation
have been so far restricted to a treatment of discrete eigenstates.
An attempt has been made in~\cite{araujo} to extend this approach to
the potentials with a mixed (discrete and continuous) eigenvalue spectrum.
However, we have found a basic error in this treatment, indicated explicitly
in the end of section~\ref{sec:l1}.

The aim of our work is to show how the problem
with mixed eigenvalue spectrum can be treated correctly.
We will show this in two examples: one with constant Schr\"{o}dinger
potential and another with P\"{o}schl-Teller potential.
The same example has been incorrectly treated in~\cite{araujo}.
To avoid any confusion one has to note that the P\"{o}schl-Teller potential is
referred to as Rosen-Morse potential in~\cite{araujo}.

\section{Solution of FPE with constant diffusion coefficient}
\label{sec:sol}

We start our consideration with the one-dimensional Fokker-Planck equation~(\ref{eq:genfp})
in the following formulation
\begin{equation}
 \frac{\partial p(x,t) }{\partial t} = -\frac{\partial}{\partial x} \left[ f(x) p(x,t) \right]
 + \frac{D}{2} \frac{\partial^2 p(x,t)}{\partial x^2} \;\label{eq:eq}
\end{equation}
for the probability density distribution $p(x,t)$, depending on the variable $x$ and time $t$.
Here, $f(x)$ is the nonlinear force and $D$ is the diffusion coefficient, which is now assumed
to be constant. We consider natural boundary conditions
\begin{equation}
 \lim\limits_{x\rightarrow\pm\infty} p(x,t)=\lim\limits_{x\rightarrow\pm\infty}
 \frac{\partial p(x,t)}{\partial x}= 0
\label{eq:bc}
\end{equation}
and take the most frequently used initial condition
\begin{equation}
p(x, t=0)=\delta(x-x_0) \label{eq:init}
\end{equation}
in the form of the $\delta$-function.
This FPE~(\ref{eq:eq}) can be transformed into an equation of Schr\"odinger type (see section~\ref{sec:tds}).
Unfortunately, the well known relation [see equation~(\ref{eq:res_tffpe})],
derived for the discrete eigenvalue spectrum,
cannot be applied if this equation has a continuous or mixed eigenvalue spectrum.
To overcome this problem, we follow a properly corrected treatment of~\cite{araujo}.
Namely, we solve the FPE with boundaries
located at $x=\pm L/2$ and then take the limit $L\rightarrow\infty$ (see section~\ref{sec:tdsm}).
This approach is used in quantum mechanics to describe unbounded states. To keep a closer
touch with quantum mechanics, here we will use the boundary conditions $p(x = \pm L/2,t)=0$,
further referred to as absorbing boundaries.

\subsection{The stationary solution}
\label{sec:stat}

The stationary solution $p_\mathrm{st}(x)$ is the long-time limit of
$p(x,t)$ at $t \to \infty$, which follows from the equation
\begin{equation}
 0 = \frac{\rd}{\rd x} \left[ f(x) p_\mathrm{st}(x) \right]
 - \frac{D}{2} \frac{\rd^2 p_\mathrm{st}(x)}{\rd x^2} \; .
\end{equation}
The force $f(x)$ can be expressed in terms of the potential $V(x)$
via $f(x)=-\rd V(x)/\rd x$. It yields
\begin{equation}
 0 = - \frac{\rd}{\rd x} \left[ \frac{\rd V(x)}{\rd x} p_\mathrm{st}(x)
 + \frac{D}{2} \frac{\rd p_\mathrm{st}(x)}{\rd x} \right] \; .
\end{equation}
Due to the natural boundary conditions, we have zero flux
\begin{equation}
 j_\mathrm{st}(x) \equiv - \frac{\rd V(x)}{\rd x} p_\mathrm{st}(x)
 - \frac{D}{2} \frac{\rd p_\mathrm{st}(x)}{\rd x} = C \qquad \text{with} \qquad C=0 \; .
\end{equation}
Thus, we have
\begin{align}
 \frac{\rd p_\mathrm{st}(x)}{\rd x} &= - \frac{2}{D} \frac{\rd V(x)}{\rd x} p_\mathrm{st}(x)\;, \\
 \frac{\rd p_\mathrm{st}(x)}{p_\mathrm{st}(x)} &= - \frac{2}{D} {\rd V(x)} \;,
\end{align}
which yields the stationary solution
\begin{equation}
 p_\mathrm{st}(x) = \mathcal{N}^{-1} Y(x) \;,
\end{equation}
where
\begin{equation}
 Y(x) \equiv \exp\left[ - \frac{2}{D} V(x) \right]
\end{equation}
has the meaning of an unnormalized stationary solution only in case of natural boundaries and
$\mathcal{N}$ is the normalization constant
\begin{equation}
 \mathcal{N} = \int_{- \infty}^{+ \infty} \rd x \exp\left[ - \frac{2}{D} V(x) \right] \; .
\label{eq:norma}
\end{equation}
This function $Y(x)$ is further used to construct a time-dependent solution.

\subsection{The time-dependent solution with discrete eigenvalues}
\label{sec:tds}

Here, we derive a time-dependent solution,
starting with the transformation $p(x,t) \to q(x,t)$ defined by
\begin{equation}
 p(x,t) = Y^{1/2}(x) \; q(x,t)
\equiv \exp\left[- \frac{2}{D} \frac{V(x)}{2}\right] q(x,t) \; .
\label{eq:trans}
\end{equation}
This transformation removes the first derivative in the original Fokker-Planck equation and
generates the equation of Schr\"odinger type for the function $q(x,t)$, i.~e.,
\begin{equation} \label{eq_12}
 \frac{\partial q(x,t)}{\partial t}
 = - V_\mathrm{S}(x) q(x,t) + \frac{D}{2} \frac{\partial^2 q(x,t)}{\partial x^2} \;,
\end{equation}
where
\begin{equation}
 V_\mathrm{S}(x) = - \left\{\frac{1}{2} \frac{\rd^2 V(x)}{\rd x^2}
- \frac{2}{D} \left[\frac{1}{2} \frac{\rd V(x)}{\rd x}\right]^2 \right\}
\end{equation}
is the so-called Schr\"odinger potential.
In the case of discrete eigenvalues, we apply the superposition ansatz
\begin{equation} \label{eq_16}
 q(x,t) = \sum_{n=0}^\infty a_n(t) \psi_n(x) \,.
\end{equation}
After inserting \eqref{eq_16} into \eqref{eq_12}, we get
the eigenvalue problem
\begin{equation}
\frac{D}{2} \, \frac{\rd^2 \psi_n(x)}{\rd x^2} - V_\mathrm{S}(x) \psi_n(x) = - \lambda_n \psi_n(x)
\label{eq:eigenvp}
\end{equation}
for eigenfunctions $\psi_n(x)$ and eigenvalues $\lambda_n \geqslant 0$
with time-dependent coefficients $a_n(t)$ given by
\begin{equation}
a_n(t) = a_n(0) \exp\left( - \lambda_n \, t \right) \; .
\end{equation}
According to this, equation~(\ref{eq_16}) can be written as
\begin{equation}
 q(x,t) = \sum_{n=0}^\infty a_n(0) \re^{-\lambda_n \, t} \psi_n(x) \;.
\label{eq:qxt}
\end{equation}
The eigenfunctions $\psi_n(x)$ are orthonormal, i.~e.,
\begin{equation}
 \int_{-\infty}^{+\infty} \psi_n(x) \psi_m(x) \rd x = \delta_{nm}
\label{eq:orthonorm}
\end{equation}
and satisfy the closure condition (completeness relation)
\begin{equation}
 \sum_{n=0}^\infty \psi_n(x') \psi_n(x) = \delta(x-x') \;.
\label{eq:completeness}
\end{equation}
Equation~(\ref{eq:eigenvp}) can be written as a
Schr\"odinger-type eigenvalue equation with Hermitian
Hamilton operator $\mathcal{H}$:
\begin{equation}
\mathcal{H} \psi_n(x) = \lambda_n \psi_n(x) \qquad \text{with} \qquad
\mathcal{H} = - \frac{D}{2} \frac{\rd^2 }{\rd x^2} + V_\mathrm{S}(x) \;.
\label{eq:H}
\end{equation}
The coefficients $a_n(0)$ in~(\ref{eq:qxt}) are calculated using the
initial condition
\begin{equation}
 p(x,t=0) = Y^{1/2}(x) q(x,t=0) = \delta(x-x_0) \;.
\end{equation}
According to~(\ref{eq:qxt}), this relation can be written as
\begin{equation}
  Y^{-1/2}(x) \delta(x-x_0) = \sum_{m=0}^\infty a_m(0) \psi_m(x) \;.
\end{equation}
In the following, we multiply both sides of this equation by $\psi_n(x)$ and integrate
over $x$ from $-\infty$ to $+\infty$. Taking into account~(\ref{eq:orthonorm}),
it yields the so far unknown coefficients
\begin{equation}
 a_n(0) = Y^{-1/2}(x_0) \psi_n(x_0) \; .
\end{equation}
The final result of this calculation reads
\begin{equation}
 p(x,t) = \sqrt{\frac{Y(x)}{Y(x_0)}}
\sum_{n=0}^\infty \re^{-\lambda_{n} \, t} \psi_{n}(x_0) \psi_{n}(x) \;.
\label{eq:res_tffpe}
\end{equation}
Note that this method can also be used for other boundary conditions.
The solution in the general form of~(\ref{eq:res_tffpe}) is well known
from older studies, e. g.,~\cite{Barrett} and can be found in many
textbooks, e.~g.,~\cite{Risken,Gardiner}.

\subsection{The time-dependent solution with mixed eigenvalue spectrum}
\label{sec:tdsm}

\begin{figure}[thbp]
 \centering
 \includegraphics[width=0.50\textwidth]{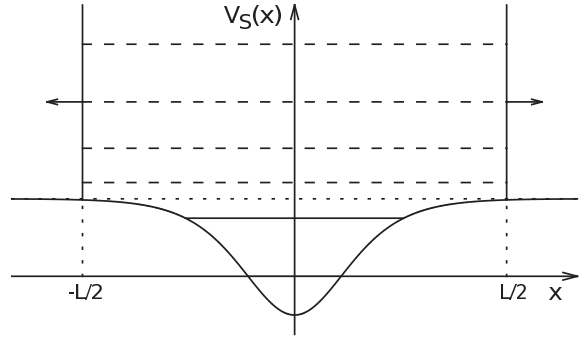}
\caption{A schematic view of the eigenvalue spectrum for the problem with two
absorbing boundaries at $x = \pm L/2$. The Schr\"odinger potential $V_S(x)$ together
with the boundaries at $x = \pm L/2$ is indicated by a solid curve and vertical lines.}
\label{fig:schematic}
\end{figure}

Consider now the problem with two absorbing boundaries
located at $x=\pm L/2$ instead of the natural boundary conditions.
In this case, we have a discrete eigenvalue spectrum, and equation~(\ref{eq:res_tffpe}) can be used
(with summation over exclusively those eigenfunctions which satisfy the boundary conditions
in a box of length $L$) to calculate
the probability distribution $p_{L}(x,t)$, i.~e.,
\begin{equation}
 p_{L}(x,t) = \sqrt{\frac{Y(x)}{Y(x_0)}}
\sum_{n=0}^\infty \re^{-\lambda_{n,L} \, t} \psi_{n,L}(x_0) \psi_{n,L}(x) \;,
\label{eq:res_tffpe1}
\end{equation}
where $\lambda_{n,L}$ are eigenvalues and $\psi_{n,L}(x)$ are the corresponding eigenfunctions,
which fulfill the boundary conditions. Let us split this infinite sum into two parts:
for $\lambda_{n,L}<\lambda_\mathrm{con}$ and $\lambda_{n,L}\geqslant\lambda_\mathrm{con}$,
where $\lambda_\mathrm{con}$ is the smallest continuum eigenvalue in the case of natural boundaries.
This eigenvalue spectrum is shown schematically in figure~\ref{fig:schematic},
where the value of $\lambda_\mathrm{con}$ is shown by a horizontal dotted line,
the eigenvalues $\lambda_{n,L}<\lambda_\mathrm{con}$~--- by solid lines and the eigenvalues
$\lambda_{n,L}\geqslant\lambda_\mathrm{con}$~--- by dashed lines.
Let $M(L)$ be the maximal value of $n$ for which $\lambda_{n,L}<\lambda_\mathrm{con}$ and
$k_{n-M(L),L}=\left[{2}(\lambda_{n,L}-\lambda_\mathrm{con})/{D}\right]^{1/2}$ for $n>M(L)$ and
$\psi^\mathrm{con}_{k_{n-M(L),L}}(x) = \psi_{n,L}(x)$ for $n>M(L)$. Hence, we have
\begin{eqnarray}
 p_{L}(x,t) &=& \sqrt{\frac{Y(x)}{Y(x_0)}}
\sum_{n=0}^{M(L)} \re^{-\lambda_{n,L} \, t} \psi_{n,L}(x_0) \psi_{n,L}(x)\nonumber \\
&&+\sqrt{\frac{Y(x)}{Y(x_0)}}\re^{-\lambda_\mathrm{con} \, t}
\sum_{m=1}^{\infty} \re^{-\frac{1}{2}D k^2_{m, L} \, t} \psi^\mathrm{con}_{k_{m, L}}(x_0) \psi^\mathrm{con}_{k_{m, L}}(x)
\;.
\label{eq:res_tffpe2}
\end{eqnarray}

The solution with natural boundaries is the limit case $L\rightarrow\infty$
\begin{equation}
 p(x,t)=\lim\limits_{L\rightarrow\infty}p_{L}(x,t)
\end{equation}
or
\begin{eqnarray}
 p(x,t) &=& \sqrt{\frac{Y(x)}{Y(x_0)}}
\sum_{n=0}^{N-1} \re^{-\lambda_{n} \, t} \psi_{n}(x_0) \psi_{n}(x)\nonumber \\
&&+\sqrt{\frac{Y(x)}{Y(x_0)}}\re^{-\lambda_\mathrm{con} \, t}
\lim\limits_{L\rightarrow\infty} \sum_{m=1}^{\infty} \re^{-\frac{1}{2} Dk^2_{m, L} \, t}
\psi^\mathrm{con}_{k_{m, L}}(x_0) \psi^\mathrm{con}_{k_{m, L}}(x)
\;,
\label{eq:res_tffpe3}
\end{eqnarray}
where $N=\lim_{L\rightarrow\infty}M(L)$ is the number of bounded states in the case with natural boundaries.
Since the eigenfunctions cannot be normalized at $L\rightarrow\infty$, it is appropriate to
write equation~(\ref{eq:res_tffpe3}) for unnormalized eigenfunctions $\bar{\psi}^\mathrm{con}_{k_{m, L}}(x)$,
\begin{eqnarray}
 p(x,t) &=& \sqrt{\frac{Y(x)}{Y(x_0)}}
\sum_{n=0}^{N-1} \re^{-\lambda_{n} \, t} \psi_{n}(x_0) \psi_{n}(x) \nonumber \\
&&
+\sqrt{\frac{Y(x)}{Y(x_0)}}\re^{-\lambda_\mathrm{con} \, t} 
\lim\limits_{L\rightarrow\infty} \sum_{m=1}^{\infty} \re^{-\frac{1}{2}D k^2_{m, L} \, t}
\underbrace{\frac{\mathcal{N}^{-1}}{\Delta k_L}}_{g^{-1}(k, L)}
\bar{\psi}^\mathrm{con}_{k_{m, L}}(x_0) \bar{\psi}^\mathrm{con}_{k_{m, L}}(x) \Delta k_L
\;,
\label{eq:un2}
\end{eqnarray}
where the normalization constant $\mathcal{N}$ is given by
 \begin{equation}
\mathcal{N}=\int_{-L/2}^{L/2}\!\!\mathrm{d}x\, |\bar{\psi}^\mathrm{con}_{k}(x)|^2
\end{equation}
and the expression under infinite sum is divided and multiplied by $\Delta k_{L}=k_{m+1, L}-k_{m, L}$.

The infinite sum can be split into two parts: one with odd $m$ and the other with even $m$.
If the Schr\"odinger potential is symmetric, then one of these two parts
contains only odd eigenfunctions $\bar{\psi}^\mathrm{o}_{k}(x)$, whereas the other
part has only even eigenfunctions $\bar{\psi}^\mathrm{e}_{k}(x)$.
In the limit $L \to \infty$, these two sums can be represented by corresponding integrals,
yielding
\begin{eqnarray}
 p(x,t) &=& \sqrt{\frac{Y(x)}{Y(x_0)}}
\sum_{n=0}^{N-1} \re^{-\lambda_{n} \, t} \psi_{n}(x_0) \psi_{n}(x)\nonumber \\
&&+\sqrt{\frac{Y(x)}{Y(x_0)}}\re^{-\lambda_\mathrm{con} \, t}
 \int_{0}^{\infty}\!\!\mathrm{d}k\, \re^{-\frac{1}{2} Dk^2 \, t} g^{-1}_\mathrm{e}(k) \bar{\psi}^\mathrm{e}_{k}(x_0) \bar{\psi}^\mathrm{e}_{k}(x)\nonumber\\
&&+\sqrt{\frac{Y(x)}{Y(x_0)}}\re^{-\lambda_\mathrm{con} \, t}
 \int_{0}^{\infty}\!\!\mathrm{d}k\, \re^{-\frac{1}{2} Dk^2 \, t} g^{-1}_\mathrm{o}(k) \bar{\psi}^\mathrm{o}_{k}(x_0) \bar{\psi}^\mathrm{o}_{k}(x)
\;,
\label{eq:res_tffpe5}
\end{eqnarray}
where
\begin{equation}
 g_\mathrm{e}(k)=\lim\limits_{L\rightarrow\infty} \left[2 \Delta k_{L} \int_{-L/2}^{L/2}\!\!\mathrm{d}x\, |\bar{\psi}^\mathrm{e}_{k}(x)|^2 \right]\,,
\label{eq:ge}
\end{equation}
\begin{equation}
 g_\mathrm{o}(k)=\lim\limits_{L\rightarrow\infty} \left[ 2 \Delta k_{L} \int_{-L/2}^{L/2}\!\!\mathrm{d}x\, |\bar{\psi}^\mathrm{o}_{k}(x)|^2 \right]\,.
\label{eq:go}
\end{equation}
This representation is useful if the eigenvalues and eigenfunctions are known.

\section{The analytical solution of FPE with constant force}

Let us consider a constant force term. In this case, the Fokker-Planck equation~(\ref{eq:eq}) reads
\begin{align}
 &\frac{\partial p(x,t) }{\partial t} = -v_{\mathrm{drift}}\frac{\partial p(x,t) } {\partial x}
 + \frac{D}{2} \frac{\partial^2 p(x,t)}{\partial x^2} \,\label{eq:cf}\,.
\end{align}
This is a drift-diffusion problem for the potential $V(x)=-v_{\mathrm{drift}}x\;$ normalized to $V(x=0)=0$.
No stationary solution exists for this problem, because the normalization constant $\mathcal{N}$
in equation~(\ref{eq:norma}) diverges in this case. Nevertheless, the transformation~(\ref{eq:trans})
$p(x,t)={Y(x)}^{1/2} q(x,t)$ with
\begin{equation}
 Y(x) = \exp\left[- \frac{2}{D} \frac{V(x)}{2}\right]
= \exp\left[ \frac{v_{\mathrm{drift}}}{D} x \right]
\end{equation}
can be used here to obtain an equation of
Schr\"{o}dinger type~(\ref{eq_12}) with constant Schr\"{o}dinger potential
\begin{equation}
 V_\mathrm{S}=\frac{1}{2D}v_{\mathrm{drift}}^2 \;.
\label{eq:constpot}
\end{equation}
The  stationary Schr\"odinger-type equation corresponding to~(\ref{eq:H}) reads
\begin{equation}
\frac{\rd^2 \psi_n(x)}{\rd x^2}  - \left[\frac{v_{\mathrm{drift}}^2}{D^2} - \frac{2}{D}\lambda_n \right] \psi_n(x) = 0 \,.
\label{eq:drift_sch}
\end{equation}

Let us now add two absorbing boundaries located at $x=\pm L/2$, where $\psi(x= \pm L/2)=0$.

Only in the case of real $k_n= \left[{2}\lambda_n/{D} - {v_{\mathrm{drift}}^2}/{D^2}\right]^{1/2}>0$ equation~\eqref{eq:drift_sch} has
non-trivial solutions
\begin{equation}
 \psi_n(x)=A \, \cos(k_nx)+ B \sin(k_nx)\,,
\end{equation}
which satisfy the boundary conditions. These solutions are
\begin{equation}
\psi_{n, L }(x)=\begin{cases}
           \sqrt{\frac{2}{L}}\cos{\left( k_{n, L} x \right)} &\qquad\text{ if $n$ is even},\\
           \sqrt{\frac{2}{L}}\sin{\left( k_{n, L} x \right)} &\qquad\text{ if $n$ is odd},
          \end{cases}
\label{eq:drift_psi}
\end{equation}
where $n = 0,1,2, \ldots$ and
\begin{equation}
k_{n,L}=\frac{\pi}{L} (n+1)\,.
\label{eq:drift_k}
\end{equation}
According to~\eqref{eq:drift_psi}--\eqref{eq:drift_k}, we have
from~(\ref{eq:ge}) and~(\ref{eq:go})
\begin{equation}
 g_\mathrm{e}(k)=g_\mathrm{o}(k)=\pi\,.
\end{equation}
Taking into account that
\begin{equation}
 \lambda_\mathrm{con} = \lim\limits_{L \to \infty} \min \{ \lambda_{n,L} \} =
\lim\limits_{L \to \infty} \min \left\{ \frac{D}{2} k_{n,L}^2 + \frac{v_{\mathrm{drift}}^2}{2D} \right\}
= \frac{v_{\mathrm{drift}}^2}{2D}
\end{equation}
holds, we obtain from equation~\eqref{eq:res_tffpe5} the expression
\begin{eqnarray}
 p(x,t)&=&\exp\left[{\frac{1}{D}v_{\mathrm{drift}}(x-x_0)}\right]\exp\left[{-\frac{v_{\mathrm{drift}}^2}{2D} \, t}\right]\nonumber \\
&&\times \frac{1}{\pi} \int_0^\infty \!\!\mathrm{d}k\re^{-\frac{1}{2}Dk^2 t}\big[\cos(kx)\cos(kx_0)+\sin(kx)\sin(kx_0)\big].
\label{eq:drift_f}
\end{eqnarray}
Using the well known identities
\begin{equation}
 \cos(kx)\cos(kx_0)+\sin(kx)\sin(kx_0)=\cos[k(x-x_0)]
\end{equation}
and
\begin{equation}
 \int_0^\infty \!\!\mathrm{d}k\; \re^{-\alpha k^2}\,\cos(\beta k)
=\sqrt{\frac{\pi}{4\alpha}}\re^{-{\beta^2}/{4\alpha}} \;,
\end{equation}
after simplification we obtain  the well known result
\begin{equation}
 p(x,t)=\frac{1}{\sqrt{2Dt}}\exp\left[-\frac{(x-x_0-v_{\mathrm{drift}}t)^2}{2Dt}\right] \;,
\label{eq:Gaussprof}
\end{equation}
which describes a moving and broadening Gaussian profile.

\section{Fokker-Planck dynamics with P\"{o}schl-Teller potential}

Here,  as a particular example  we consider the force
\begin{equation}
 f(x)=-b\tanh \left(\alpha x\right)
\end{equation}
with some positive constants $b$ and $\alpha$. This corresponds to the diffusion problem in
the potential
\begin{equation}
 V(x)=\frac{b}{\alpha}\ln\left(\cosh \alpha x\right)\,,
\label{eq:pot}
\end{equation}
normalized to $V(x=0)=0$.
\begin{figure}[!b]
 \centering
 \includegraphics[width=0.50\textwidth]{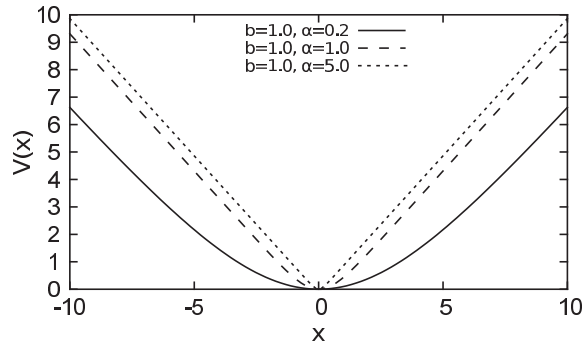}
\caption{Graphical representation of equation~(\ref{eq:pot}) for $b=1$ and several values of parameter $\alpha$.}
\label{fig:vx}
\end{figure}

\begin{figure}[!t]
 \begin{center}
 \includegraphics[width=0.50\textwidth]{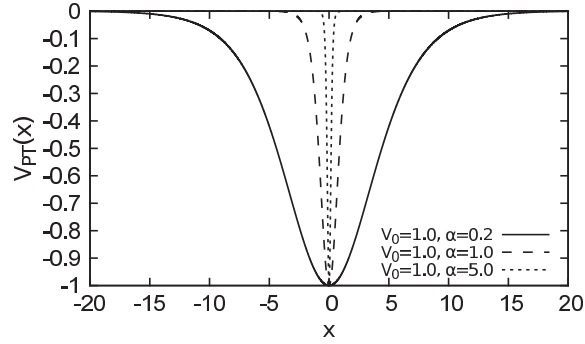}
 \end{center}
 \vspace{-3mm}
\caption{ P\"oschl-Teller potential (\ref{eq:posch}) for $V_0=1$ and several values of the parameter $\alpha$.}
\label{fig:posch}
\end{figure}

Figure~\ref{fig:vx} shows that this potential is actually a smoothed version of the V-shaped potential.
The corresponding Schr\"{o}dinger potential in this case is
\begin{equation}
 V_\mathrm{S}(x)=\frac{b^2}{2D} -\left(\frac{b^2}{2D}+\frac{b\alpha}{2}\right)
\frac{1}{\cosh^2(\alpha x)} \;.
\label{eq:V_S}
\end{equation}
If we compare it [see  equation~(\ref{eq:posch}) and figure~\ref{fig:posch}] with
the well known P\"oschl-Teller potential
\begin{equation}
 V_\mathrm{PT}(x)= V_\mathrm{S}(x) - \frac{b^2}{2D} = -\frac{V_0}{\cosh^2\left(\alpha x\right)} \;,
\label{eq:posch}
\end{equation}
we see that equation~(\ref{eq:V_S}) represents the shifted by ${b^2}/{2D}$
P\"oschl-Teller potential with $V_0={b^2}/{2D}+{b\alpha}/{2}$.
As we can see from figure~\ref{fig:posch}, the P\"oschl-Teller potential gives a mixed
(discrete and continuous) eigenvalue spectrum. Therefore, equation~(\ref{eq:res_tffpe})
cannot be directly used to solve the FPE. We have to use~(\ref{eq:res_tffpe5}).

The eigenvalue equation~(\ref{eq:eigenvp}) for the potential~(\ref{eq:V_S}) reads
\begin{equation}
 \frac{D}{2}\frac{\rd^2\psi_n(x)}{\rd x^2} - \left[\frac{b^2}{2D} -\left(\frac{b^2}{2D}
+\frac{b\alpha}{2}\right) \frac{1}{\cosh^2(\alpha x)} \right]\psi_n(x)= -\lambda_n\psi_n(x) \;.
\label{eq:schrod}
\end{equation}
By introducing dimensionless variables $\tilde{x}=\alpha x$, $\tilde{l}={b}/{D\alpha}$ and
$\tilde{\lambda}_n={2\lambda_n}/{D\alpha^2}-\tilde{l}^{2}$, we write (\ref{eq:schrod}) in a dimensionless form
\begin{equation}
 -\frac{\rd^2\psi_n(\tilde{x})}{\rd\tilde{x}^2}-\tilde{l}\left(\tilde{l}+1\right)
\frac{1}{\cosh^2 \tilde{x}} \psi_n(\tilde{x})=\tilde{\lambda}_n\psi_n(\tilde{x})\, .
\label{eq:schrod_no_dim}
\end{equation}
Analytical solutions for both bounded and unbounded eigenfunctions of equation~(\ref{eq:schrod_no_dim})
are known and can be found in~\cite{Nieto,Lekner}.

\subsection{Bounded solutions for P\"oschl-Teller potential}

 The equation~(\ref{eq:schrod_no_dim}) has $N=\max\,\{m\in\mathbb{N}\mid m< \tilde{l}+1\}\,$  bounded states
$n=0,1,2, \ldots, N-1$, where $\mathbb{N}$  is a set of all natural numbers $ \mathbb{N} = \{ 0, 1, 2, \ldots \}$.
Here, we consider the eigenfunctions
with $\tilde{\lambda}_n=0$ as unbounded, because they cannot be normalized.

The eigenvalues can be calculated from the following equation~\cite{Nieto}
\begin{equation}
 \tilde{\lambda}_n=-(\tilde{l}-n)^2 \;, \qquad \text{for }\qquad n<N\,; \quad  n\in\mathbb{N}\,.
\end{equation}
Note that at least one bounded state with $\tilde{\lambda}_0=-\tilde{l}^2$ always exists for $\tilde{l}>0$, which corresponds
to $\lambda_0=0$.
The bounded eigenfunctions are known~\cite{Nieto}
\begin{equation}
 \psi_n\left(\tilde{x}\right)=\cosh^{-\tilde{l}}(\tilde{x}) \times
 \begin{cases} \mathcal{N}_\mathrm{e}(n)\,\mathrm{F}\big(-\frac{1}{2}n,\frac{1}{2}n-\tilde{l};\frac{1}{2};
-\sinh^2\tilde{x} \big) &\qquad{\text{ if } \ n \ \text{ is even}}, \\
 \mathcal{N}_\mathrm{o}(n)\sinh (\tilde{x}) \,\mathrm{F}\bigl(\frac{1}{2}-\frac{n}{2},\frac{n}{2}+\frac{1}{2}-\tilde{l};
\frac{3}{2};-\sinh^2\tilde{x} \bigr) &\qquad{\text{ if } \ n \ \text{ is odd}}, \end{cases}
\label{eq:psin}
\end{equation}
where $\mathrm{F}$ denotes a hypergeometric function, which can be represented by Gaussian hypergeometric series
\begin{equation}
 \mathrm{F}(\alpha, \beta; \gamma;\zeta)=\frac{\mathrm{\Gamma}(\gamma)}{\mathrm{\Gamma}(\alpha)\mathrm{\Gamma}(\beta)}
\sum\limits_{k=0}^{\infty}\frac{\mathrm{\Gamma}(\alpha+k)\mathrm{\Gamma}(\beta+k)}{\mathrm{\Gamma}(\gamma+k)}\frac{\zeta^n}{n!} \;.
\end{equation}
The normalization constants are
\begin{align}
& \mathcal{N}_\mathrm{e}(n)=\left[\frac{2\left(\tilde{l}-n\right)}{\left(\tilde{l}-\frac{1}{2}n\right)
 \left(n+1\right) }
\frac{1}{\mathrm{B}\big(\frac{1}{2}, \tilde{l}-\frac{1}{2}n\big)\mathrm{B}\big(\frac{1}{2}, 1+\frac{1}{2}n\big)}\right]^{1/2}\;,\\
%
%
&\mathcal{N}_\mathrm{o}(n)=\left[\frac{2\left(\tilde{l}-n\right)}{\tilde{l}-\frac{1}{2}\left(n+1\right) }
\frac{1}{\mathrm{B}\big(\frac{3}{2}, \tilde{l}-\frac{1}{2}(n+1)\big)\mathrm{B}\big(\frac{1}{2}, \frac{1}{2}(n+1)\big)}\right]^{1/2}\;,
\end{align}
where $\mathrm{B}(a, b)$ is the beta function $\mathrm{B}(a, b)=\mathrm{\Gamma}(a)\mathrm{\Gamma}(b)/\mathrm{\Gamma}(a+b)$.

\subsection{Unbounded solutions for P\"oschl-Teller potential}

The unbounded solutions have a continuous eigenvalue spectrum with $0\leqslant\tilde{\lambda}<\infty$.
Thus, we can introduce $\tilde{k}={\tilde{\lambda}}^{1/2}$ (with $\tilde{k}=k/\alpha$).
The P\"{o}schl-Teller potential is symmetric.
Therefore, the eigenfunctions are the even  and odd functions known from~\cite{Lekner}
\begin{align}
\bar{\psi}_{\tilde{k},\tilde{l}}\left(\tilde{x} \right)&= A\cdot\psi^\mathrm{e}_{\tilde{k},\tilde{l}}\left(\tilde{x} \right)+B\cdot \psi^\mathrm{o}_{\tilde{k},\tilde{l}}\left(\tilde{x} \right)\,, \\
\bar{\psi}^\mathrm{e}_{\tilde{k},\tilde{l}}\left(\tilde{x} \right)&=
 \left(\cosh\tilde{x} \right)^{\tilde{l}+1} \mathrm{F}\Big(r,s;\frac{1}{2};-\sinh^2\tilde{x} \Big)\,, \label{eq:evenf} \\
\psi^\mathrm{o}_{\tilde{k},\tilde{l}}\left(\tilde{x} \right)&=
 \left(\cosh\tilde{x} \right)^{\tilde{l}+1} \sinh( \tilde{x}) \, \mathrm{F}\Big(r+\frac{1}{2},s+\frac{1}{2};\frac{3}{2};
-\sinh^2\tilde{x} \Big)\,, \label{eq:oddf}
\end{align}
 where $A$ and $B$ are constants, and
\begin{equation}
 r=\frac{1}{2}\left(\tilde{l}+1+\ri\tilde{k}\right)\,,\quad s=\frac{1}{2}\left(\tilde{l}+1-\ri\tilde{k}\right)\,.
\end{equation}
Since these are unbounded solutions, eigenfunctions cannot be normalized within $x\in(-\infty; +\infty)$.

As we see, the eigenfunctions are rather complicated in general case.
The expressions become essentially simpler
for integer values of $\tilde{l}$. Therefore, without loosing the general idea, we will show the solutions of the
Fokker-Planck equation for $\tilde{l}=1$ and
$\tilde{l}=2$.

\subsection{The solution of FPE for P\"oschl-Teller potential with parameter
$\tilde{l}=1$}
\label{sec:l1}

For $\tilde{l}=1$ (which implies $b=\alpha D$) we have only one bounded state with
the eigenvalue $\tilde{\lambda}_0=-1$ and the eigenfunction [equation~(\ref{eq:psin}) for $n=0$]
\begin{equation}
 \psi_0(\tilde{x})=\frac{1}{\sqrt{2}\cosh(\tilde{x})}\,.
\label{eq:psi0}
\end{equation}
 The unbounded eigenfunctions~(\ref{eq:evenf}) and~(\ref{eq:oddf}) are
\begin{align}
\bar{\psi}^\mathrm{e}_{\tilde{k}}\left(\tilde{x} \right)&=
\cos(\tilde{k}\tilde{x})-\frac{1}{\tilde{k}}\tanh(\tilde{x})\sin(\tilde{k}\tilde{x})\,,\\
\bar{\psi}^\mathrm{o}_{\tilde{k}}\left(\tilde{x} \right)&=
\sin(\tilde{k}\tilde{x})+\frac{1}{\tilde{k}}\tanh(\tilde{x})\cos(\tilde{k}\tilde{x})\,.
\end{align}

As proposed in section~\ref{sec:sol}, we add two absorbing boundaries located at
$\tilde{x}=\pm \tilde{L}/2$. Due to these boundary conditions, we have only discrete values of $\tilde{k}$.
Let us denote them by $\tilde{k}_{\tilde{L},m}$ for even functions and by $\tilde{\kappa}_{\tilde{L},m}$
for odd functions. The values of $\tilde{k}_{\tilde{L},m}$ and  $\tilde{\kappa}_{\tilde{L},m}$, obtained
from the boundary conditions, are positive solutions of the transcendent equations
\begin{equation}
 \tilde{k}_{\tilde{L},m}=\tanh({\tilde{L}}/{2})\tan(\tilde{k}_{\tilde{L},m}\tilde{L}/{2}),
\label{eq:nl1}
\end{equation}
\begin{equation}
 \tilde{\kappa}_{\tilde{L},m} \tan(\tilde{\kappa}_{\tilde{L},m})=-\tanh({\tilde{L}}/{2}),
\label{eq:nl2}
\end{equation}
where $m=1,2,3,\ldots$ denotes the $m$-th smallest positive solution.
The equations for normalized eigenfunctions now read as
\begin{align}
\psi^\mathrm{e}_{\tilde{k}_{\tilde{L},m}}\left(\tilde{x} \right)&
=\mathcal{N}^{-1/2}_\mathrm{e}(\tilde{k}_{\tilde{L},m}, \tilde{L})\cdot\left[
\cos(\tilde{k}_{\tilde{L},m} \tilde{x})-
\frac{1}{\tilde{k}_{\tilde{L},m}}   \tanh(\tilde{x})  \sin(\tilde{k}_{\tilde{L},m} \tilde{x})\right]\,,\\
\psi^0_{\tilde{\kappa}_{\tilde{L},m}}\left(\tilde{x} \right)&
= \mathcal{N}^{-1/2}_\mathrm{o}(\tilde{\kappa}_{\tilde{L},m}, \tilde{L})\cdot\left[
\sin(\tilde{\kappa}_{\tilde{L},m}\tilde{x})
+\frac{1}{\tilde{\kappa}_{\tilde{L},m}}   \tanh(\tilde{x})  \cos(\tilde{\kappa}_{\tilde{L},m}\tilde{x})\right]\,,
\end{align}
where normalization constants for odd and even eigenfunctions are
\begin{equation}
 \mathcal{N}_\mathrm{e}(\tilde{k}, \tilde{L}) =
\frac{\left(\tilde{k}^2+1\right)
\left[\tilde{k} \tilde{L} - \sin(\tilde{k}\tilde{L}) \right] }{2\tilde{k}^3}
\,,
\label{eq:nl3}
\end{equation}
\begin{equation}
 \mathcal{N}_\mathrm{o}(\tilde{k}, \tilde{L}) =
\frac{\left(\tilde{k}^2+1\right)
\left[\tilde{k} \tilde{L}   +  \sin(\tilde{k}\tilde{L}) \right] }{2\tilde{k}^3}
\,.
\label{eq:nl4}
\end{equation}

In the limit case $\tilde{L}\rightarrow\infty$, equations~(\ref{eq:nl1})--(\ref{eq:nl2})
for the allowed $\tilde{k}$ values, as well as equations~(\ref{eq:nl3})--(\ref{eq:nl4})
for the normalization constants simplify to
\begin{eqnarray}
&&  \tilde{k}_{\tilde{L}\rightarrow\infty,m} =\frac{2m\pi}{\tilde{L}}\,,
\hspace{1cm} \Delta \tilde{k}_{\tilde{L}\rightarrow\infty} = \frac{2\pi}{L} \,,\\
&& \tilde{\kappa}_{\tilde{L}\rightarrow\infty,m} =\frac{(2m-1)\pi}{\tilde{L}}\,,
\quad \Delta \tilde{\kappa}_{\tilde{L}\rightarrow\infty} = \frac{2\pi}{L} \,,\\
&&\mathcal{N}_\mathrm{e}(\tilde{k}, \tilde{L}\rightarrow\infty) = \mathcal{N}_\mathrm{o}(\tilde{k}, \tilde{L}\rightarrow\infty)
=\frac{L}{2}\frac{\tilde{k}^2+1}{\tilde{k}^2} \,,
\end{eqnarray}
and we  also have
\begin{align}
g_\mathrm{e}(\tilde{k}) = \Delta \tilde{k}_{\tilde{L}\rightarrow\infty} \cdot \mathcal{N}_\mathrm{e}(\tilde{k}, \tilde{L}\rightarrow\infty)
  =\pi\frac{\tilde{k}^2+1}{\tilde{k}^2} \;, \\
  g_\mathrm{o}(\tilde{\kappa}) = \Delta \tilde{\kappa}_{\tilde{L}\rightarrow\infty} \cdot \mathcal{N}_\mathrm{o}(\tilde{\kappa}, \tilde{L}\rightarrow\infty)
  =\pi \frac{\tilde{\kappa}^2+1}{\tilde{\kappa}^2}\,.
\end{align}
Inserting these relations as well as $\lambda_\mathrm{con}=\tilde{l}^2 \alpha^2 D/2$
(following from $\tilde{\lambda}_\mathrm{con}= {2 \lambda_\mathrm{con} }/{D \alpha^2 }- \tilde{l}^2=0$)
into~(\ref{eq:res_tffpe5}), we finally obtain the time-dependent solution of the
Fokker-Planck equation
\begin{eqnarray}
 p(x,t)&=&\frac{1}{2\cosh^2(\alpha x)}\nonumber
\\[1ex]&&+
\frac{\cosh(\alpha x_0)}{\pi \cosh(\alpha x)} \re^{-\frac{1}{2}D\alpha^2 t} \int_0^\infty \!\!\mathrm{d}\tilde{k} \;\re^{-\frac{1}{2}D\alpha^2 \tilde{k}^2 t} \frac{\tilde{k}^2}{\tilde{k}^2+1} \bar{\psi}^\mathrm{e}_{\tilde{k}}\left(\alpha x \right) \bar{\psi}^\mathrm{e}_{\tilde{k}}\left(\alpha x_0 \right)\nonumber
\\&&+
\frac{\cosh(\alpha x_0)}{\pi \cosh(\alpha x)} \re^{-\frac{1}{2}D\alpha^2 t}
\int_0^\infty \!\!\mathrm{d} \tilde{k} \;\re^{-\frac{1}{2}D\alpha^2 \tilde{k}^2 t}
\frac{\tilde{k}^2}{\tilde{k}^2+1} \bar{\psi}^\mathrm{o}_{\tilde{k}}\left(\alpha x \right) \bar{\psi}^\mathrm{o}_{\tilde{k}}\left(\alpha x_0 \right) \;.
\label{eq:solutionl1}
\end{eqnarray}
If the initial condition is given by $x_0=0$, then $\psi^\mathrm{o}_{\tilde{k}}\left(0 \right)=0$ and
$\psi^\mathrm{e}_{\tilde{k}}\left(0 \right)=1$ hold, which allows us to obtain a simpler expression
\begin{eqnarray}
 p(x,t)&=&\frac{1}{2\cosh^2(\alpha x)} +
\frac{1}{\pi \cosh(\alpha x)} \re^{-\frac{1}{2}D\alpha^2  t}\nonumber\\
 &&\times \int_0^\infty \!\!\mathrm{d}\tilde{k} \;\re^{-\frac{1}{2}D\alpha^2 \tilde{k}^2 t} \frac{\tilde{k}^2}{\tilde{k}^2+1}
\left[\cos(\tilde{k}\alpha x)-\frac{1}{\tilde{k}}\tanh(\alpha x )\sin(\tilde{k}\alpha x) \right]\,.
\end{eqnarray}

\begin{figure}[!t]
 \centering
 \includegraphics[width=0.50\textwidth]{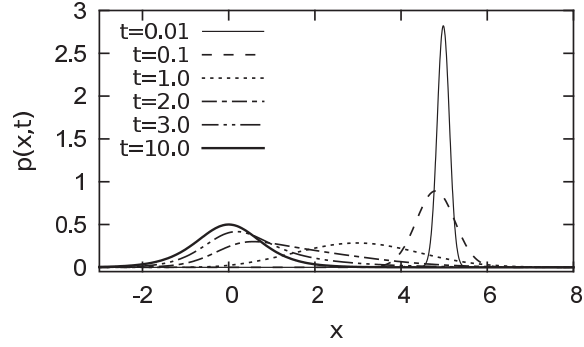}
\caption{The probability distribution at different time moments $t$, calculated for the parameters
$b=2$, $D=2$ and $\alpha=1$ ($\tilde{l}=1$) starting at $x_0=5$.}
\label{fig:l1}
\end{figure}

The solution for parameters $b=2$, $D=2$ and $\alpha=1$, corresponding to $\tilde{l}=1$,
with the initial location of the delta-peak at $x_0=5$ is shown
in figure~\ref{fig:l1} for different time moments $t$.
As we can see, the probability distribution moves to the left. It broadens at
the beginning. For larger times, it becomes narrower again and converges to the stationary
solution $p_\mathrm{st}(x)= \lim_{t \to \infty} p(x,t) = [\cosh^2(\alpha x)]^{-1} =
{\psi_0(x)}^2$ [see equations~(\ref{eq:solutionl1}) and~(\ref{eq:psi0})], which is a symmetric distribution around
$x=0$. The stationary solution is
practically reached at $t=10$. This behavior is expected from the drift-diffusion dynamics.

For small times $t \to 0$, we have a delta-peak located at $x=x_0$ in accordance
with the given initial condition~(\ref{eq:init}). For comparison, the ``general solution''
of~\cite{araujo} does not satisfy this initial condition due to a wrong construction,
where the contribution of bounded states is simply summed up with a Gaussian probability
density profile (calculated with an error). The latter corresponds to unbounded states for zero Schr\"odinger
potential at $L \to \infty$, as it is evident from~(\ref{eq:Gaussprof}) and~(\ref{eq:constpot}) at
$v_{\mathrm{drift}}=0$. Therefore, the result appears to be correct only at $t \to \infty$
when the Gaussian part vanishes.
It is clear that the whole set of eigenfunctions should be calculated self-consistently for the
given potential to obtain a correct and meaningful result, since only in this case the completeness
relation~(\ref{eq:completeness}) holds and all different eigenfunctions are orthogonal.
Thus, the basic error of~\cite{araujo} is that some of the eigenfunctions are calculated
for zero Schr\"odinger potential in~\cite{araujo}, whereas all of them should be calculated for
the true Schr\"odinger potential.

\subsection{The solution of FPE for P\"oschl-Teller potential with parameter
$\tilde{l}=2$}

For $\tilde{l}=2$ (which implies $b=2\alpha D$) we have two bounded states with
eigenvalues $\tilde{\lambda}_0=-4$ and $\tilde{\lambda}_1=-1$. The corresponding eigenfunctions are
\begin{eqnarray}
&& \psi_0(\tilde{x})=\frac{\sqrt{3}}{2\cosh^2(\tilde{x})}\,,\\
&& \psi_1(\tilde{x})=\sqrt{\frac{3}{2}} \, \frac{\sinh(\tilde{x})}{\cosh^2(\tilde{x})}\,.
\end{eqnarray}
The unbounded eigenfunctions are
\begin{align}
\bar{\psi}^\mathrm{e}_{\tilde{k}}\left(\tilde{x} \right)&=
\left[1+\tilde{k}^2 -3\tanh^2(\tilde{x}) \right]\cos(\tilde{k}\tilde{x})-3\tilde{k} \tanh(\tilde{x})\sin(\tilde{k}\tilde{x})\,,\\
\bar{\psi}^\mathrm{o}_{\tilde{k}}\left(\tilde{x} \right)&=
\left[1+\tilde{k}^2 -3\tanh^2(\tilde{x}) \right]\sin(\tilde{k}\tilde{x}) +3\tilde{k} \tanh(\tilde{x})\cos(\tilde{k}\tilde{x})\,.
\end{align}

By adding again two absorbing boundaries at
$\tilde{x}=\pm \tilde{L}/2$, we have discrete values of $\tilde{k}$, i.~e.,
$\tilde{k}_{\tilde{L},m}$ for even functions
and $\tilde{\kappa}_{\tilde{L},m}$ for odd functions.
In the limit $\tilde{L}\rightarrow\infty$, we  again obtain
the classical infinite-square-well relations for eigenstates:
\begin{eqnarray}
&&  \tilde{k}_{\tilde{L}\rightarrow\infty,m} =\frac{(2m-1)\pi}{\tilde{L}}\,,\\
&& \tilde{\kappa}_{\tilde{L}\rightarrow\infty,m} =\frac{2m\pi}{\tilde{L}}\,.
\end{eqnarray}
The normalization constants in this case are
\begin{equation}
\mathcal{N}_\mathrm{e}\left(\tilde{k}, \tilde{L}\rightarrow\infty\right) = \mathcal{N}_\mathrm{o}\left(\tilde{k}, \tilde{L}\rightarrow\infty\right)
=\frac{L}{2} {\left(\tilde{k}^2+4\right)\left(\tilde{k}^2+1\right)}\, .
\end{equation}

By applying the same steps as in the case of $\tilde{l}=1$, we obtain the solution
\begin{eqnarray}
 p(x,t)&=&\frac{3}{4\cosh^4(\alpha x)}
+ \frac{3}{2} \frac{\sinh(\alpha x)\sinh(\alpha x_0)}{\cosh^4(\alpha x)} \re^{-\frac{3}{2}D\alpha^2 t}\nonumber
\\[1ex]&&+
\frac{\cosh^2(\alpha x_0)}{\pi \cosh^2(\alpha x)} \re^{-2D\alpha^2 t} \int_0^\infty \!\!\mathrm{d}\tilde{k} \;\re^{-\frac{1}{2}D\alpha^2 \tilde{k}^2 t}\frac{1}{\tilde{k}^2+5\tilde{k}^2+4} \psi^\mathrm{e}_{\tilde{k}}\left(\alpha x \right) \psi^\mathrm{e}_{\tilde{k}}\left(\alpha x_0 \right)\nonumber
\\&&+
\frac{\cosh^2(\alpha x_0)}{\pi \cosh^2(\alpha x)} \re^{-2 D\alpha^2 t} \int_0^\infty \!\!\mathrm{d} \tilde{k} \;\re^{-\frac{1}{2}D\alpha^2 \tilde{k}^2 t} \frac{1}{\tilde{k}^2+5\tilde{k}^2+4} \psi^\mathrm{o}_{\tilde{k}}\left(\alpha x \right) \psi^\mathrm{o}_{\tilde{k}}\left(\alpha x_0 \right).
\end{eqnarray}

\begin{figure}[!t]
 \centering
 \includegraphics[width=0.50\textwidth]{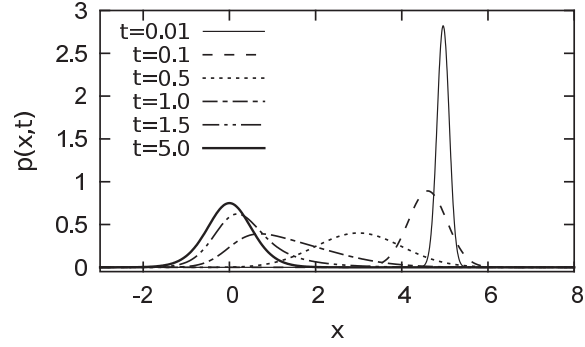}
\caption{The probability distribution at different time moments $t$, calculated for the parameters
$b=2$, $D=4$ and $\alpha=1$ ($\tilde{l}=2$) starting at $x_0=5$.}
\label{fig:l2}
\end{figure}

The solution for parameters $b=4$, $D=2$ and $\alpha=1$, corresponding to $\tilde{l}=2$,
with the initial condition given by $x_0=5$ is shown
in figure~\ref{fig:l2} for different time moments $t$.
The evolution of the probability distribution is very similar to
that one shown in figure~\ref{fig:l1} for $\tilde{l}=1$, with the only
essential difference that the dynamics is faster and the distribution
is somewhat narrower due to a deeper potential well.

\section{Conclusions}

Using the analogy of the Fokker-Planck equation with the Schr\"odinger equation,
it has been shown how the time-dependent solution can be constructed in the case
of mixed eigenvalue spectrum with free and bounded states.
The method is based on the idea of introducing two absorbing boundaries
at $x = \pm L/2$, considering the limit $L \to \infty$ afterwards.
Although this idea is similar to the one proposed earlier in~\cite{araujo},
it is obvious that the problem is quite non-trivial, so that the oversimplified
(i.e., erroneous) approach of~\cite{araujo} cannot be used~--- see discussion in the end of section~\ref{sec:l1}.
Analytical solutions have been found and analyzed in two examples of the Schr\"odinger
potential being constant (constant force) and a shifted P\"{o}schl-Teller potential.
For the latter potential, the analytical solutions have been compared with the
results of the Crank-Nicolson numerical integration method, and the agreement within
an error of $10^{-7}$ has been found.
The time evolution of the calculated probability distribution in these examples is consistent
with the usual drift-diffusion dynamics.

\section*{Acknowledgement}

The authors M. B. and J.~K. thank for financial support from Academic Exchange Office at Rostock University having made it possible to continue our long-standing collaboration between Rostock (Germany) and Riga (Latvia).


%
%

\ukrainianpart

\title{Як розв'язати рівняння Фоккера-Планка, використовуючи спектр змішаних власних значень?%
}
\author{М. Брицс\refaddr{label1}, Я. Каупузс\refaddr{label2},
Р. Манке\refaddr{label1}}
\addresses{
\addr{label1}  Інститут фізики, Університет м. Росток, D--18051 Росток, Німеччина
\addr{label2}  Інститут математики і комп'ютерних наук, Латвійський
університет,   LV--1459 Рига, Латвія}

\makeukrtitle

\begin{abstract}
\tolerance=3000
Аналогія рівняння Фоккера-Планка  (FPE) з рівнянням Шредингера
дозволяє використати метод квантової механіки для знаходження
аналітичного розв'язку  FPE для низки випадків. Проте, попередні
дослідження обмежувалися потенціалом Шредингера з дискретним
спектром власних значень. Тут ми покажемо, як цей підхід можна також
застосувати до спектру змішаних власних значень зі зв'язаними і
вільними станами. Ми розв'язуємо  FPE з границями, що знаходяться
при $x=\pm L/2$ і беремо границю  $L\rightarrow\infty$, розглядаючи
приклади з постійним потенціалом Шредингера і потанціалом
Пешля-Теллера. Спрощений підхід  раніше запропонували  M.T.~Араухо та
E.~Дріго Фільйо. Детальне дослідження двох прикладів показує, що
коректний розв'язок, отриманий в цій статті, узгоджується з
очікуваною динамікою Фоккера-Планка.

\keywords рівняння Фоккера-Планка, рівняння Шредингера, потенціал
Пешля-Теллера
\end{abstract}

\lastpage
\end{document}